\documentstyle[pre,aps,multicol,epsf]{revtex}

\begin{document}   

\draft				     

\title{Spiral Waves in Media with Complex Excitable Dynamics}
\author{Andrei Goryachev and Raymond Kapral}
\address{Chemical Physics Theory Group, Department of
Chemistry, University of Toronto, Toronto, ON M5S 3H6, Canada}
\maketitle
\begin{abstract}
The structure of spiral waves is investigated in super-excitable 
reaction-diffusion systems where the local dynamics exhibits multi-looped 
phase space trajectories. It is shown that such systems support stable
spiral waves with broken rotational symmetry and complex temporal  
dynamics. The main structural features of such waves, synchronization 
defect lines, are demonstrated to be similar to those of spiral waves in
systems with complex-oscillatory dynamics.
\end{abstract}


\begin{multicols}{2} 

\narrowtext 

Studies of spatially-distributed active
media have demonstrated the ubiquity of self-organized spatio-temporal patterns,
in particular spiral waves, in various physical or biological systems such 
as the Belousov-Zhabotinsky (BZ) reaction \cite{BZ}, catalytic surfaces
\cite{cat}, cardiac muscle \cite{card} and colonies of the amoebae {\it Dictiostelium
discoideum} \cite{DD}. 

Most research has been devoted to the study of simple oscillatory 
or excitable systems. Recently it was shown that reactive media with 
complex periodic and chaotic oscillations are capable of supporting spiral
waves \cite{CP-gen} with a variety of distinctive features, absent 
in simple oscillatory systems \cite{CP}. The rotational symmetry of 
spiral waves in period-doubled media is broken by synchronization 
line defects where the phase of the local oscillation changes by multiples 
of $2\pi$. It was conjectured that spiral waves with broken rotational 
symmetry could also be observed in super-excitable systems where the phase
space trajectory after excitation follows a multi-looped 
path of relaxation to the stable fixed point \cite{SEx}. Broken spirals 
with a clearly visible synchronization defect line emanating from the 
spiral core were observed under special three-dimensional conditions 
in the BZ reactive medium \cite{jap}. The nature of spiral waves in 
complex-excitable media is nevertheless largely unexplored.

In this paper we show that spiral waves with broken rotational symmetry 
exist in a prototypical super-excitable system and demonstrate that 
the topological properties of line defects, described earlier for 
oscillatory media, also hold for excitable systems.

\begin{figure}[htbp]
\begin{center}
\leavevmode
\epsffile{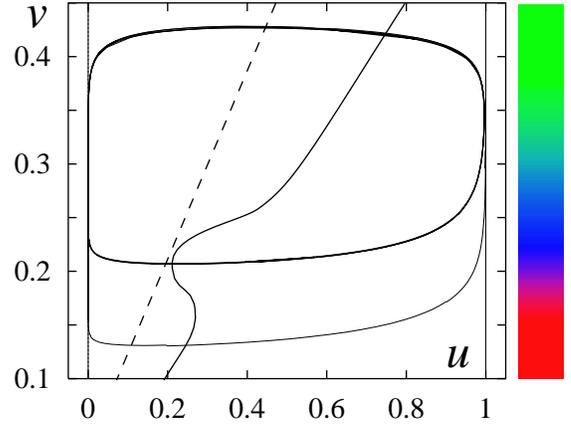}
\end{center}
\caption{Phase plane $(u,v)$ plot of a ${\bf 1^2}$ type trajectory  
calculated at a point in the medium shown in Fig.\ref{p3}. Two stable 
branches $(u=0, u=1)$ and one unstable branch of the nullcline $\dot{u}=0$ 
are shown as solid lines and the nullcline $\dot{v}=0$ by a dashed line.}
\label{phpo}
\end{figure}

We consider a spatially-distributed system whose dynamics is governed by a 
pair of reaction-diffusion equations of the form
\begin{eqnarray}
\label{rds}
{\partial  u \over \partial t} & = & \frac{1}{\varepsilon}u(1-u)(u-\frac{v+b}{a} -
f(v)) + D \nabla^2  u \;, \\	 \nonumber
{\partial  v \over \partial t} & = & u-v \;,
\end{eqnarray}
where $v$ is a non-diffusive variable and both $u$ and $v$ are functions of 
time and space. This model with $f(v)\equiv0$ was studied in \cite{bark} 
as a simplified version of the FitzHugh-Nagumo model which serves as a 
prototype of an excitable system described by two variables.
The excitable dynamics of system (\ref{rds})  consists of 
two fast and two slow stages. If displaced from the stable fixed point 
$u=0,v=0$ to the right of the unstable branch of the nullcline 
$\dot{u}=0$, it quickly reaches upper stable branch $u=1$. It follows 
this branch until $v$ reaches sufficiently large values and then jumps
to the lower stable branch $u=0$, along which it slowly relaxes to the 
stationary state. To add a super-excitability to (\ref{rds}) a 
modification of the unstable branch of the
nullcline $\dot{u}=0$ was proposed in \cite{arr} as
\begin{equation}
\label{mod}
f(v)=\alpha\exp\left(-\frac{(v-v_0)^2}{\sigma^2}\right) \; .
\end{equation}
The introduction of (\ref{mod}) changes the shape of the unstable branch of
$\dot{u}=0$ so that for suitably chosen parameters $\alpha,\sigma$ and $v_0$ 
it nearly touches the nullcline $\dot{v}=0$ (see Fig.\ref{phpo}).
As a result, if another excitation is applied to the system (\ref{rds}) 
before it has reached the stationary state, it may execute second, smaller 
excitable loop before it finally reaches the stable state. 

The spatio-temporal dynamics of a one-dimensional array of such elements 
forced by an external pacemaker with varying period was studied 
in \cite{arr}. When the period of forcing $T_f$ is larger than a certain 
internal period of the system $T_0$, the response is a train of waves 
corresponding to the large relaxation loop $({\bf 1^0})$. If $T_f<T_0$ the 
system develops wavetrains with low amplitude corresponding to small 
relaxation loop $({\bf 0^1})$. Non-trivial behavior
is observed when $T_f$ is only slightly smaller than $T_0$. In this case 
the system shows mixed-mode waveforms $({\bf 1^n})$ consisting of one large 
and $n$ small waves. Response of this type is an example of the 
complex-excitable dynamics targeted in our studies. Instead of an external 
pacemaker we use a spiral wave as a self-sustained source of excitation 
in the medium.			   
\begin{figure}[htbp]
\begin{center}
\leavevmode
\epsffile{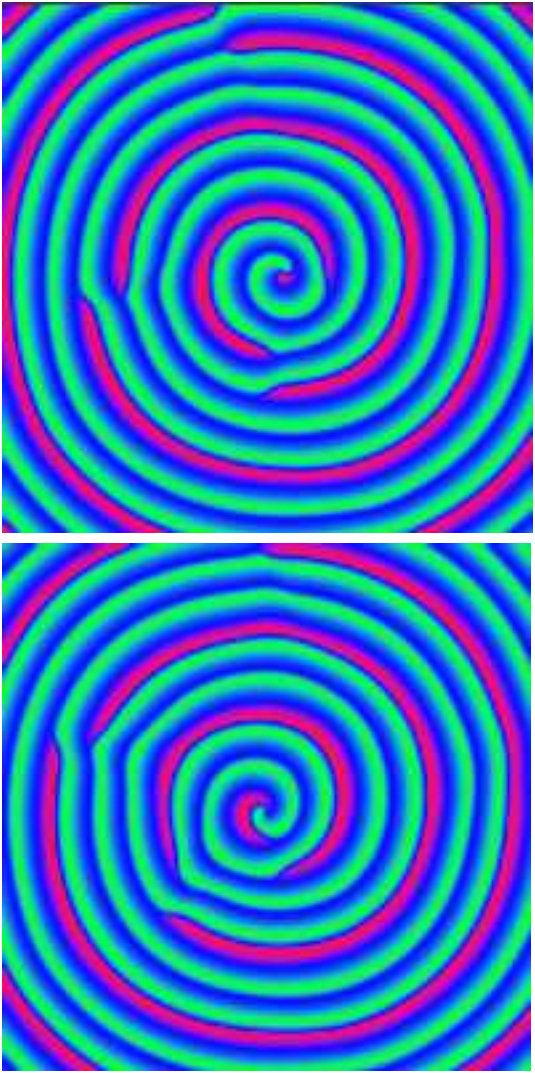}
\end{center}
\caption{ An irregular spiral wave with complex excitable dynamics is 
shown at two different times in the upper and lower panels, respectively.
The concentration field $v({\bf r},t_0)$ is color-coded as shown in 
Fig.\ref{phpo}.}
\label{mix}
\end{figure}

\begin{figure}[htbp]
\begin{center}
\leavevmode
\epsffile{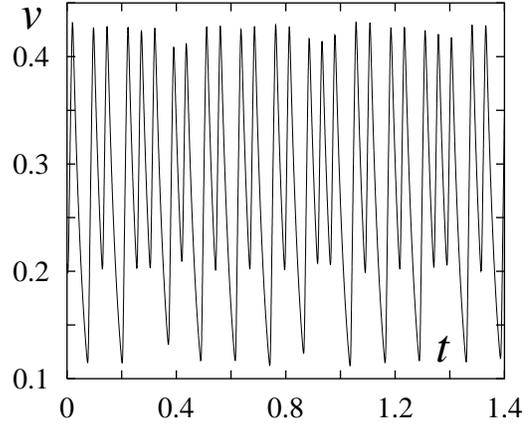}
\end{center}
\caption{Time series $v({\bf r},t)$ calculated at a point in the medium
shown in Fig.\ref{mix}.}
\label{mix_ts}
\end{figure}

Spiral waves were initiated in a two-dimensional square domain with no-flux 
boundary conditions. While $\alpha$ and $\sigma$ were used as bifurcation 
parameters, other parameters were fixed at ($\varepsilon=0.005,\,a=0.6,\, 
b=0.03,\, v_0=0.2$). As in \cite{arr} complex-excitable dynamics was found 
in the parameter region between domains of large-amplitude ${\bf 1^0}$ 
(small $\alpha$ and $\sigma$) and small-amplitude ${\bf 0^1}$ (large 
$\alpha$ and $\sigma$) waves. In this region the medium supports mostly 
aperiodic, stable spiral waves lacking rotational
symmetry. Figure \ref{mix} shows such a spiral wave for $\alpha=0.15,\;
\sigma^2=0.001$ at two time instances. Note that the wave length of a 
large-amplitude wave is larger than that of a small-amplitude wave and, thus, 
the shape of the spiral is distorted. The concentration time 
series $v({\bf r},t)$ at different 
locations in the medium (cf Fig.\ref{mix_ts}) show aperiodic concatenations 
of ${\bf 1^1}$ and ${\bf 1^2}$ waveforms, while trivial patterns 
${\bf 1^0}$ and ${\bf 0^1}$ are completely absent.

As the domain of complex-excitable dynamics is traversed from 
${\bf 1^0}$ to ${\bf 0^1}$ the contribution of the waveforms ${\bf 1^n}$ 
with $n>0$ steadily grows, as well as the number $n$ of the small-amplitude 
loops. Thus, the transition from large to small-amplitude spiral waves occurs 
gradually through a succession of irregular spiral 
patterns with a progressively growing contribution of low-amplitude waves.

\begin{figure}[htbp]
\begin{center}
\leavevmode
\epsffile{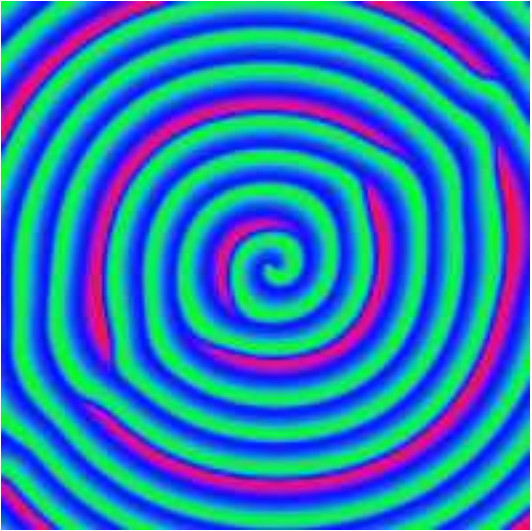}
\end{center}
\caption{ Spiral wave with period-3 dynamics. Color-coding is the same as in 
Fig.\ref{mix}.}
\label{p3}
\end{figure}

Although irregular patterns are found in most of the complex-excitable 
domain, pure period-3 dynamics was found in a sub-domain of this region. 
Figure \ref{p3} shows the spatial structure of a spiral wave in this 
parameter region. Observation of the spiral wave dynamics for long 
time periods shows that the entire concentration field slowly rotates 
around the spiral core with a constant angular velocity $\omega$. 
The period-3 dynamics is manifested in a coordinate frame centered at 
the spiral core and rotating with velocity $\omega$. Indeed, it takes 
three rotations of the spiral for the concentration field to return 
to itself. 			

Figure \ref{phpo} shows a phase portrait 
of the dynamics at a non-special location in the medium
calculated in the rotating frame. Before closing onto
itself the phase space trajectory executes two small loops and one large 
loop, corresponding to a pure ${\bf 1^2}$ dynamics. Consider a 
polar coordinate frame $(\rho,\phi)$ in the $(u,v)$ phase plane with origin 
at an arbitrary point internal to both the small and large loops. 
During one full period of the dynamics the phase variable
$\phi$ changes by $6\pi$. Calculation of the phase at every point
in the medium at a time $t_0$ gives an instantaneous snapshot 
$\phi({\bf r},t_0)$ of the time-dependent phase field $\phi({\bf r},t)$.

Consider a closed contour $\Gamma$ that surrounds the spiral core. 
The phase increment $\Delta_\Gamma
\phi=\oint_\Gamma \nabla \phi({\bf r},t_0)d{\bf l}$ along $\Gamma$ 
will be equal to a multiple of the full $6\pi$ period of the dynamics. 
~From the topological theory of point defects \cite{top} in simple 
oscillatory or excitable media it follows that $\Delta_\Gamma \phi$ 
is invariant and for the one-armed spiral waves in this study takes 
the values $\pm 2\pi$. In complex-periodic media this contradiction
is resolved by the existence of synchronization defect lines emanating 
from the core \cite{CP}. The phase of the local oscillation experiences
jumps equal to multiples of $2\pi$ when such line defects are crossed. 
Any contour $\Gamma$ encircling the spiral core 
intersects these lines so that the total phase increment 
$\Delta_\Gamma \phi$ is obtained from the integration of 
$\nabla \phi({\bf r},t_0)$ along $\Gamma$ yielding $\pm 2\pi$ 
plus phase jumps at the intersections of $\Gamma$ with the synchronization 
defect lines. The sum of both contributions yields the full period phase 
increment of the local dynamics. It was predicted \cite{CP}
that one should be able to find phenomena analogous to synchronization 
defect lines in media with complex-excitable dynamics.

For the specific case of period-3 complex-excitable dynamics where the 
total phase increment is $6\pi$, one might expect to find one line defect 
where the phase jumps by $4\pi$ or two defect lines where the phase jumps
by $2\pi$ on crossing each line. In Fig.\ref{p3} one sees two lines 
emanating from the spiral core at an angle of $\sim 180^o$. 
Investigation of the change in phase of the local dynamics across 
these lines shows that they are indeed $2\pi$ synchronization defect
lines exhibiting the loop exchange phenomenon described earlier for 
complex-oscillatory media.

\begin{figure}[htbp]
\begin{center}
\leavevmode
\epsffile{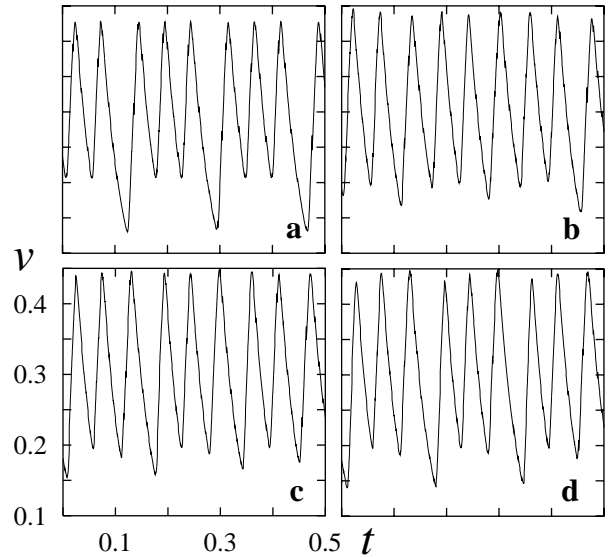}
\end{center}
\caption{Change in character of local dynamics across one of the defect 
lines seen in Fig.\ref{p3}. See text for details.}
\label{lex}
\end{figure}

Figure \ref{lex} shows the $v({\bf r},t)$ time series at four 
neighboring locations in the medium along a path traversing one of the
defect lines. Panel ${\bf a}$ is a plot of the normal ${\bf 1^2}$ 
dynamics seen on one side of the defect line. Every third minimum is 
lower than the two preceding minima and corresponds
to the larger relaxation loop in the phase space plot.
As one approaches 
the defect line, the large-amplitude loop shrinks while both small-amplitude 
loops grow (Fig.\ref{lex}(b)). Then, one small-amplitude loop begins 
to grow faster than the other and at some point becomes 
larger than the still shrinking large-amplitude loop (Fig.\ref{lex}(c)). 
This loop exchange process continues until the new large loop attains 
a size equal to that of a large-amplitude loop and the other two 
loops shrink to the size of the small-amplitude loop. The local
dynamics on opposite sides of the defect line (compare Figs \ref{lex}(a) 
and \ref{lex}(c)) experiences a $2\pi$ phase shift. Thus, the total phase 
increment $\Delta_\Gamma \phi=2\pi$ resulting from the integration of 
$\nabla \phi({\bf r},t_0)$ along $\Gamma$, excluding its intersections
with defect lines, plus the two $2\pi$ phase jumps at the 
intersection points gives the expected $6\pi$ phase increment.

Our results show that the structure of complex-periodic spiral waves 
is governed by general topological principles independent of whether the 
dynamics is excitable or oscillatory. This fact allows one to extend the 
predictions inferred from the studies of complex-oscillatory
systems to systems with super-excitable dynamics. The formation of 
complex-periodic spiral waves in such systems might play a role in  
the development of some pathological conditions in the heart. Indeed, 
mixed-mode electrical activity with alternating large and small 
amplitude maxima, so-called alternans, is typically observed as a 
symptom of tachycardia \cite{alt}.

\end{multicols}

\end{document}